\documentclass{svproc}
\usepackage{graphicx}
\usepackage[sort, numbers]{natbib}
\usepackage{url}

\begin{document}
\mainmatter              
\title{Direct \emph{N-body} code on low-power embedded ARM GPUs}
\titlerunning{Direct $N$-body on ARM SoC}  
%
\author{David Goz\inst{1} \and Sara Bertocco
\and Luca Tornatore \and Giuliano Taffoni}
\authorrunning{David Goz et al.} 
%
\tocauthor{David Goz, Sara Bertocco, Luca tornatore, and Giuliano Taffoni}
\institute{INAF - OATs, via Tiepolo 11, Trieste, Italy,\\
\email{david.goz@inaf.it},\\ WWW home page:
\texttt{http://www.oats.inaf.it}}

\maketitle              

\begin{abstract}
This work arises on the environment of the \texttt{ExaNeSt} project aiming at design and development of an exascale ready supercomputer with low energy consumption profile but able to support the most demanding scientific and technical applications. The \texttt{ExaNeSt} compute unit consists of densely-packed low-power 64-bit ARM processors, embedded within Xilinx FPGA SoCs. SoC boards are heterogeneous architecture where computing power is supplied both by CPUs and GPUs, and are emerging as a possible low-power and low-cost alternative to clusters based on traditional CPUs.

A state-of-the-art direct $N$-body code suitable for astrophysical simulations has been re-engineered in order to exploit SoC heterogeneous platforms based on ARM CPUs and embedded GPUs. 
Performance tests show that embedded GPUs can be effectively used to accelerate real-life scientific calculations, and that are promising also because of their energy efficiency, which is a crucial design in future exascale platforms.


\end{abstract}

\section{Introduction}
Nowadays ARM delivers technology to drive power-efficient System-on-Chip (hereafter SoC) solutions combining CPU and GPU into unified compute sub-system offering double-precision floating point arithmetic, and options for high performance I/O and memory interface. Those systems represent an excellent solution to build less expensive and more power-efficient computational clusters than standard High Performance Computing (HPC) facilities.

In the last years, some effort has been devoted to investigate the potential of SoCs for computationally intensive real-life scientific applications, comparing the performances with those obtained on a typical x86 HPC node (e.g \cite{Morganti_2016}). They conclude that considering SoCs for computationally intensive scientific applications seems very promising, and this technology might represent the next revolution in high performance community. 
However, scientific applications are usually ported on those platforms rather than to be re-engineered in order to fully exploit the new hardware by solving the architecture-application performance gap, i.e. the gap between the capabilities of the hardware (HW) and the performance released by the HPC software (SW).
This is crucial when designing the new generation of HPC supercomputer, the Exascale platform. The realization of an Exascale supercomputer requires significant advances in a variety of technologies, the aim of which is the energy efficiency.
A number of projects has been financed in Europe to develop an exascale-class prototype system (e.g. ExaNeSt project\footnote{http://www.exanest.eu}, Montblanc project\footnote{http://montblanc-project.eu}, and Mango project\footnote{http://www.mango-project.eu}).

The \texttt{ExaNest H2020} project \citep{ExaNeSt} aims at the design and 
development of an exascale-class prototype system built upon power-efficient hardware able to execute ambitious real-world applications coming from a wide range of scientific and industrial domains, including also HPC for astrophysics \citep{8049832}. 
Since the power-efficiency is the main concern, the ExaNeSt basic compute unit consists of low-energy-consumption ARM CPUs, embedded GPUs, FPGAs and low-latency high throughput interconnects \citep{KATEVENIS201858}.
An approach based on HW/SW co-design is crucial to design Exascale resources that can be effectively exploited by real scientific applications.

The work presented in this paper aims to study whether a direct $N$-body code, called \texttt{Hy-Nbody}, for real scientific production may benefit from embedded GPUs given that powerful high-end GPGPUs have already demonstrated to provide tremendous performance benefit.
This is the first work to implement such algorithm on embedded GPU and to compare results with multi-core on SoC implementation.
\texttt{Hy-Nbody} code is a re-engineered version of {\texttt{HiGPUs}} \citep{HiGPUs1,HiGPUs2,HiGPUs3}, a state-of-the-art direct $N$-body code, based on the Hermite 6th order time integrator, which has been widely used for scientific production in Astrophysics, i.e. for simulations of star clusters with up to $\sim$ 8 million bodies \citep{Spera1,Spera2}, and of galaxy mergers \citep{Spera3}. Moreover, \texttt{HiGPUs} has been extensively tested on large supercomputer such as IBM iDataPlex DX360M3 Linux Infiniband Cluster provided by the italian supercomputing consortium CINECA using up to 256 GPGPUs.
All kernels of the \texttt{Hy-Nbody} code have been implemented using OpenCL language\footnote{http://www.khronos.org/opencl/} in order to write efficient code for hybrid architecture.
{\texttt{Hy-Nbody}} has been designed to fully exploit computational nodes in a cluster by means of an hybrid parallelization schema, MPI+OpenMP for host code and OpenCL for device code.

This paper is organized as follows. Section \ref{section:N_body} describes direct $N$-body solvers used in Astrophysics. 
Section \ref{section:codeimpl} presents the {\texttt{Hy-Nbody}} code and its optimizations in order to exploit ARM CPU and Mali GPU.
Section \ref{section:cluster} describes the computational test bed based on Rockchip Firefly RK3399 Soc boards based on Linux Operating system.
Section \ref{section:performance} is devoted to present and discuss the results. Future development and scope of this work are presented in Section~\ref{section:development_scope}, and the conclusions in Section~\ref{section:conclusions}.

\section{N-body solvers running on hybrid computing platforms}
\label{section:N_body}
$N$-body solvers provide the backbone for different scientific and engineering applications, such as astrophysics, nuclear physics, molecular dynamics, fluid mechanics and biology. In astrophysics, the $N$-body problem is the problem of predicting the individual motions of a group of celestial objects interacting with each other only gravitationally. The classical (i.e. non relativistic) direct $N$-body problem has analytic solution only with $N=2$, so in general it must be simulated using numerical methods.  
The numerical solution of the direct $N$-body problem is still considered a challenge despite the significant advances in both hardware technologies and software development.
The main drawback related to the direct $N$-body problem relies on the fact that the algorithm requires $O(N^{2})$ computational cost. 
In practice many variations of the naive algorithm are used, for instance, implementing high order Hermite integration schemes \citep{Nitadori_2008} and block time-stepping. These variations can eliminate most of the standard parallelization method for $N^{2}$ algorithm, requiring huge effort to maximize performance.

There are some $N$-body codes designed to speed up the classical $N$-body problem for real scientific production in astrophysics using GPGPUs, for example, {\texttt{$\varphi$GPU}} code \citep{phiGPU}, {\texttt{$\varphi$GRAPE}} code \citep{Harfst_2007}, {\texttt{NBODY6}} code \citep{NBODY6}, {\texttt{MYRIAD}} code \citep{MYRIAD}, and {\texttt{HiGPUs}} code \citep{HiGPUs1,HiGPUs2,HiGPUs3}.
In all of them, the GPGPU is fed by the host CPU with the gravity equation of data in the form of coordinates, velocities and masses of particles, and it handles calculating the forces for the data points.

None of the above has been optimized and ported on embedded GPUs, which is the scope tackled by this paper.

\section{Code implementation}
\label{section:codeimpl}

\begin{figure}[!ht]
\centering\includegraphics[width=\linewidth]{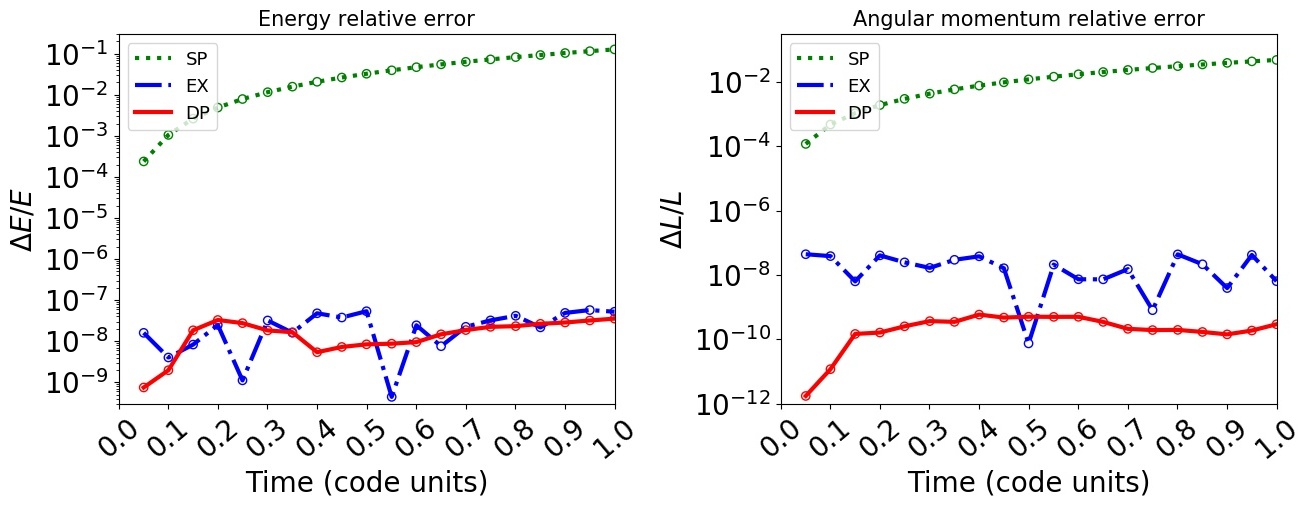}
\caption{Left panel: relative error $\Delta E/E$ for DP-arithmetic (continuous red line), EX-arithmetic (dot-dashed blue line) and SP-arithmetic (dotted green line) as a function of the integration time (in code unit).
Right panel: relative error $\Delta L/L$ for DP-arithmetic (continuous red line), EX-arithmetic (dot-dashed blue line) and SP-arithmetic (dotted green line) as a function of the integration time (in code unit).}
\label{fig:energy_momentum_conservation}
\end{figure}

The 6th order Hermite integrator for astrophysical $N$-body simulations consists of three stages (physical/mathematical aspects are described in \cite{Nitadori_2008}): a predictor step that predicts particle's positions and velocities; an evaluation step to evaluate new accelerations, their first order (\textit{jerk}), second order (\textit{snap}), and third order derivatives (\textit{crackle}); a corrector step that corrects the predicted positions and velocities using the results of the previous steps.

In the following we describe \texttt{Hy-Nbody}, which has been conceived to fully exploit the compute capabilities of heterogeneous architecture:

\begin{enumerate}

\item \textbf{DEVICE EXPLOITABLE}: the entire 6th order Hermite schema is implemented and optimized using OpenCL kernels, allowing to test the code on any OpenCL-compliant device (CPUs/GPUs/FPGAs);

\item \textbf{PARALLELIZATION SCHEMA}: a one-to-one correspondence between MPI processes and computational nodes is established and each MPI process manages all the OpenCL-compliant devices of the same type available per node (device type is selected by the user). Inside of each shared-memory computational node parallelization is achieved by means of OpenMP environment. Hence, in \texttt{Hy-Nbody}, the host code is parallelized with hybrid MPI+OpenMP programming, while the device code is parallelized with OpenCL. The user is allowed to choose at compile time if the application uses MPI or OpenMP, or both, or neither;

\item \textbf{DECOMPOSITION OVER HOST AND DEVICE}:
the Hermite integration is performed on the selected OpenCL-compliant device(s).
The algorithm uses a share-time step scheme that integrates all particles. Thus, in a simulation with $N$-particles using $n$ devices, during the evaluation stage each device deals with $N/n$ particles and evaluates $N (N/n)$ accelerations and their derivatives, subsequently collected and reduced from the all set of computational nodes. 
In \texttt{Hy-Nbody} the evaluation of the time step, by means of the so called \textit{generalized Aarseth criterion} \citep{Nitadori_2008}, the total energy and the angular momentum of the system are performed on the device as well. 
The latest quantities are periodically evaluated during the simulation in order to check the accuracy of the integration schema.

Theimplementation requires that particle data is communicated between the host and the device at each share-time step, which gives rise to synchronization points between host and device(s).
Accelerations, \textit{jerk}, \textit{snap} and time step computed by the device(s) are retrieved by the host on every computational node, reduced and then sent back again to the device(s);

\item \textbf{KERNEL OPTIMIZATIONS}: OpenCL kernels have access to distinct memory regions distinguished by access type and scope. \textit{Local} memory provides read-and-write access to work-items within the same work-group (OpenCL terminology) and it is specifically designed to reduce the latency of data transactions.
The \textit{evaluation} kernel is the computationally most expensive part of the Hermite algorithm. This makes the calculation of the accelerations a good candidate for exploiting the local memory of the device. When the \textit{evaluation} kernel is issued to the device(s), each work-item goes through the following steps:
\begin{description}
\item[i)] each work-item in the work-group caches one particle from \textit{global} memory into the \textit{local} memory. The total number of cached particles is therefore equal to the work-group size (selected by the user);
\item[ii)] partial acceleration, \textit{jerk} and \textit{snap} for each work-item are calculated and stored in registers using particles cached in local memory;
\item[iii)] steps i) and ii) are repeated until all particles handled by the device have been read (avoiding to sum up the self interaction);
\item[iv)] results are stored in global memory ready to be read by the host.
\end{description}
The previous schema implies $N(N/n)$ calculations performed by the device, requiring internal synchronization due to the fact that \textit{local} memory is limited. However, the exploiting of local \textit{memory} is generally accepted as the best method to reduce global memory latency in discrete GPUs;

\item \textbf{KERNEL VECTORIZATION}: since the majority of OpenCL-compliant devices supports vector instruction set, all kernels of the application have been vectorized. Vectorizing code can effectively improve memory bandwidth because of regular memory access, better coalescing of these memory accesses and reducing the number of loads/stores (each load/store is larger);

\item \textbf{PRECISION}: high precision computations are necessary for many numerical and scientific applications. Indeed, the Hermite 6th order integration schema requires double precision (DP) arithmetic in the evaluation of inter-particles distance and acceleration in order to minimize the round-off error.
Full IEEE-compliant DP-arithmetic is efficient in available CPUs and GPGPUs, but it is still extremely resource-eager and performance-poor in other accelerators like embedded GPUs or FPGAs.
As an alternative, the extended-precision (EX) numeric type can represent a trade-off in porting \texttt{Hy-Nbody} on devices not specifically designed for scientific calculations, such as embedded GPUs or FPGAs. An EX-number provides approximately 48 bits of mantissa at single-precision exponent ranges. 
\texttt{Hy-Nbody} can be run using DP, EX or single precision (SP) arithmetic (user-defined at compile time). The EX-arithmetic is implemented as proposed by \cite{EX}. 

To test the effect of the arithmetic on the accumulation of the round-off error, the energy $E$ and the angular momentum $L$ of the $N$-body system during the simulation are compared with the values at the start of the simulation.
Latest quantities must remain constant within an isolated system. The relative errors $\Delta E/ E$ and $\Delta L / L$ are determined using the following equations:
\begin{equation}
\frac{\Delta E}{E} = \frac{\left |E_{start} - E(t) \right |}{E_{start}} , \quad \textrm{and} \quad \frac{\Delta L}{L} = \frac{\left |L_{start}  -L(t) \right |}{L_{start}}
\end{equation}
where $E_{start}$, $L_{start}$ and $E(t)$, $L(t)$ are the energy and the angular momentum at the start and at a given time of the simulation, respectively.
Figure \ref{fig:energy_momentum_conservation} shows the relative errors of energy and angular momentum of the system as a function of time (in code unit). The simulation was carried out with 4096 particles. 
As relevant result, adopting SP-arithmetic, round-off error accumulates during the simulation, while it is roughly constant using EX or DP-arithmetic.
The test presented suggests that EX-arithmetic can be effectively adopted for $N$-body problem ensuring to keep control over the accumulation of the round-off error during the simulation. This approach requires only 32-bit compute capability to the computational device.

\end{enumerate}

\subsection{Tuning OpenCL kernels for the embedded ARM Mali GPU}
\label{subsection:ARM_opt}
OpenCL is a portable language but it not always performance portable, so existing OpenCL code is typically tuned for specific architecture.
However, general purpose programming for embedded GPUs is still relatively new, and the associated runtime libraries are often immature.

ARM developer guide\footnote{\url{infocenter.arm.com/help/topic/com.arm.doc.100614_0303_00_en/arm_mali_gpu_opencl_developer_guide_100614_0303_00_en.pdf}} says that for best performance on Mali-T864 (Mali Midgard family) the code should be vectorized to achieve the best performance. Regardless of the native width of the GPU's SIMD functional units, using wider vectors in the kernel may provide the GPU architecture more opportunity for exploiting data-level parallelism. Kernels in {\texttt{Hy-Nbody}} have already been vectorized, since also discrete GPUs show enhanced performances exploiting vectorization.
On the Mali GPU, moreover, the global and local OpenCL address spaces are mapped to main host memory. This means that explicit data copies from global to local memory with associated barrier synchronizations are not necessary. Thus, using local memories as a cache can waste both performance and power on the Mali GPU. A specific ARM-GPU-optimized version of all kernels of {\texttt{Hy-Nbody}} has been implemented in which the local memory is not used.

\section{Testbed description}
\label{section:cluster}

\begin{figure}[!ht]
\centering\includegraphics[width=\linewidth]{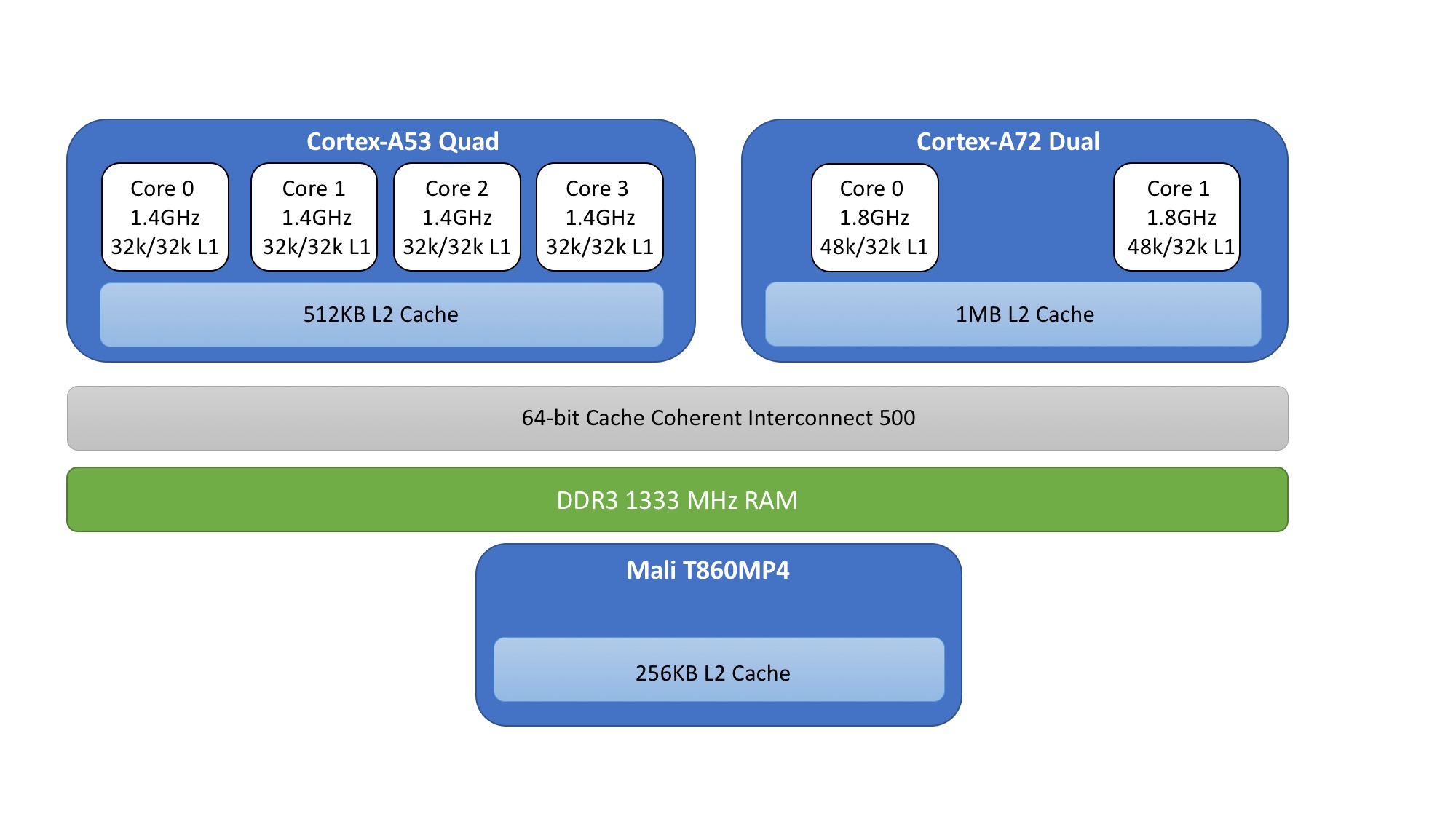}
\caption{ Rockchip Firefly-RK3399 design.}
\label{fig:Firefly-RK3399}
\end{figure}

Waiting for the ExaNest prototype release, the \texttt{Hy-Nbody} code has been validated and tested on a deployed cluster based on heterogeneous ARM-hardware (CPUs + embedded GPUs).
Each computational node is a Rockchip Firefly-RK3399 single board computer. 
It is a six core 64-bit High-Performance Platform, based on SoC with the ARM big.LITTLE architecture. ARM big.LITTLE technology features two sets of cores: a low performance energy-efficient cluster that is called “LITTLE” and power hungry high performance cluster that is called “big”. Rockchip Firefly-RK3399 SoC is presented in Figure~\ref{fig:Firefly-RK3399}. 
Each board contains (1) a cluster of four Cortex-A53  cores with 32kB  L1 cache and 512 L2 cache, and (2) a cluster of two Cortex-A72 high-performance cores with 32kB L1 cache and 1M L2 cache. Each cluster operates at independent frequencies, ranging from 200MHz up to 1.4GHz for the LITTLE and up to 1.8GHz for the big. The SoC contains 4GB DDR3 - 1333MHz  RAM. The L2 caches are connected to the main memory via the 64-bit Cache Coherent Interconnect (CCI) 500 that provides full cache coherency between big.LITTLE processor clusters and provides I/O coherency for the Mali-T864 GPU. The peculiarity of this board is that Mali-T864 is a OpenCL-compliant Quad-Core ARM Mali GPU.

The main characteristics of this cluster, named \texttt{INCAS}\footnote{\textbf{IN}tensive \textbf{C}lustered \textbf{A}rm-\textbf{Soc}}, are listed in Table~\ref{table:INCAS}.
The cluster is managed by SLURM (Simple Linux Utility for Resource Management), a free and open-source job scheduler for Linux and Unix-like kernels, used by many of the world’s supercomputers and computer clusters.

Results presented in the following section have been carried out by means of \texttt{INCAS}, which is fully described in \cite{Bertocco_INAF_report_1}.

\begin{table}
\begin{center}
\begin{tabular}{|c|c|}
\hline
Cluster name & \texttt{INCAS}\\
\hline
Nodes available & 8\\
\hline
SoC         & Rockchip RK3399 (28nm HKMG Process)\\
\hline
CPU   & Six-Core ARM 64-bit processor \\
& (Dual-Core Cortex-A72 and Quad-Core Cortex-A53)\\
\hline
GPU & ARM Mali-T864 MP4 Quad-Core GPU\\
\hline
Ram memory & 4GB Dual-Channel DDR3\\
\hline
Network & 1000Mbps Ethernet\\
\hline
Power & DC12V - 2A (per node)\\
\hline
Operating System & Ubuntu version 16.04\\
\hline
Compiler & gcc version 7.3.0\\
\hline
MPI      & OpenMPI version 3.0.1\\
\hline
OpenCL   & OpenCL 2.2\\
\hline
Job scheduler & SLURM version 17.11\\
\hline
\end{tabular}
\end{center}
\caption{The main characteristics of our cluster used to test the \texttt{Hy-Nbody} code.}
\label{table:INCAS}
\end{table}

\section{Performance results}
\label{section:performance}

The 6th order Hermite integration schema implemented in {\texttt{Hy-Nbody}} relies on three different stages, described in Section~\ref{section:codeimpl}. The \textit{evaluation} stage is the most computationally demanding, considering that with $N$-bodies the algorithm requires $O(N^{2})$ computational cost.
The performance of the \textit{evaluation} kernel is measured for both ARM CPU and GPU, testing how the running time (average of 10 runs of the kernel) changes as a function of the number of OpenMP threads in the CPU code, and of the work-group size in the GPU code. On the GPU side the impact of specific ARM-GPU-optimizations, as discussed in Section~\ref{subsection:ARM_opt}, are investigated. Performances have been measured for both DP and EX precision arithmetic.

\subsection{ARM CPUs performance results}

\begin{figure}[!ht]
\centering\includegraphics[width=\linewidth]{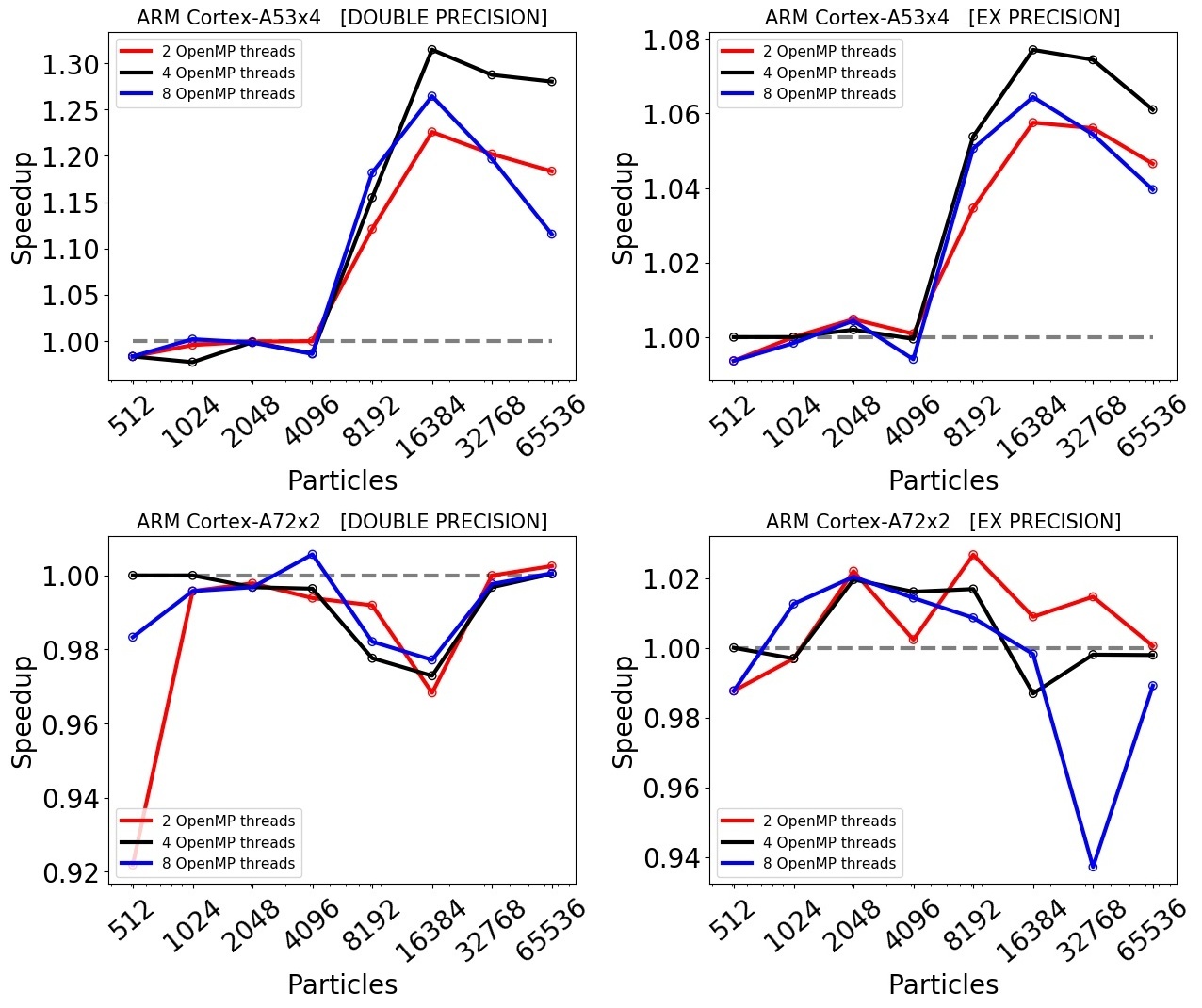}
\caption{Host speedup for DP-arithmetic (left panels) and EX-arithmetic (right panels) varying OpenMP threads as a function of the number of particles. Top panels for ARM Cortex-A53x4 CPU and bottom panels for ARM Cortex-A72x2 CPU.}
\label{fig:host_CPUs_D_EX}
\end{figure}

\begin{figure}[!ht]
\centering\includegraphics[width=\linewidth]{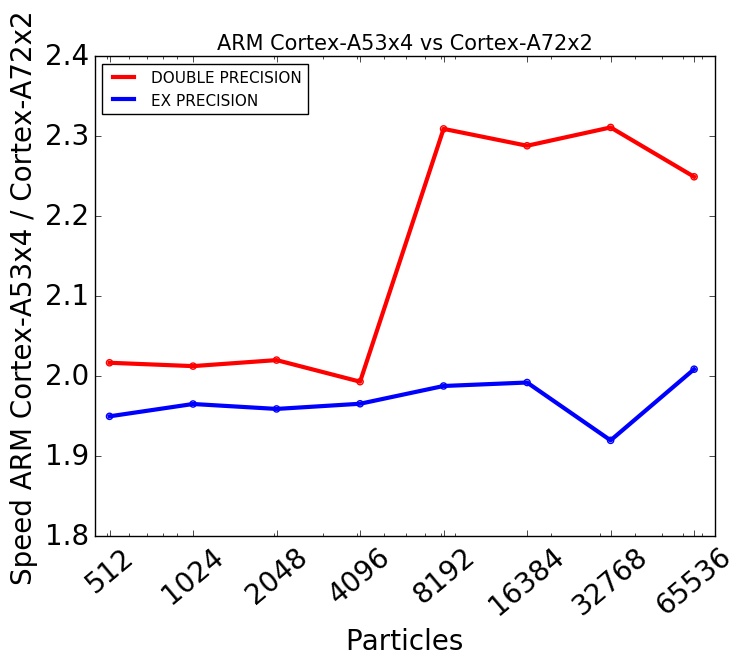}
\caption{Speed comparison between ARM Cortex-A53x4 and Cortex-A72x2 CPUs for both DP-arithmetic (red line) and EX-arithmetic (blue line).}
\label{fig:host_CPUs_comparison_time}
\end{figure}

ARM big.LITTLE processors have three main software execution models: cluster migration (a single cluster is active at a time, and migration is triggered on a given workload threshold), CPU migration (pairing every big core with a LITTLE core), and heterogeneous multiprocessing  mode (also known as Global Task Scheduling, which allows using all of the cores simultaneously).

The CPUs speedup, i.e. the ratio of the serial execution time to the parallel execution time utilizing multiple cores by means of OpenMP threads, is measured and studied. Kernel execution time on both ARM Cortex-A53x4 and Cortex-A72x2 CPUs have been obtained setting explicit CPU affinity and using the Linux system function \texttt{getrusage}, getting the total amount of time spent executing in user mode.

Figure~\ref{fig:host_CPUs_D_EX} shows the speedup for both ARM Cortex-A53x4 and Cortex-A72x2 CPUs varying the number of OpenMP threads as a function of the number of particles. On the ARM Cortex-A53x4, for both DP-arithmetic and EX-arithmetic, some speedup is obtained only when the number of particles exceeds 4096 in number. As expected, the best performance is achieved with four OpenMP threads, where most likely there is one thread per available core. On the ARM Cortex-A72x2 one thread is always faster then multiple threads for DP-arithmetic and only a minor speedup is achieved with two threads adopting EX-arithmetic.

Figure~\ref{fig:host_CPUs_comparison_time} shows the ratio of the best running time achieved by the CPUs as a function of the number of particles for both arithmetic. ARM Cortex-A72x2 is faster than Cortex-A53x4 by approximately a factor of two.

\subsection{ARM embedded GPU performance results}

\begin{figure}[!ht]
\centering\includegraphics[width=\linewidth]{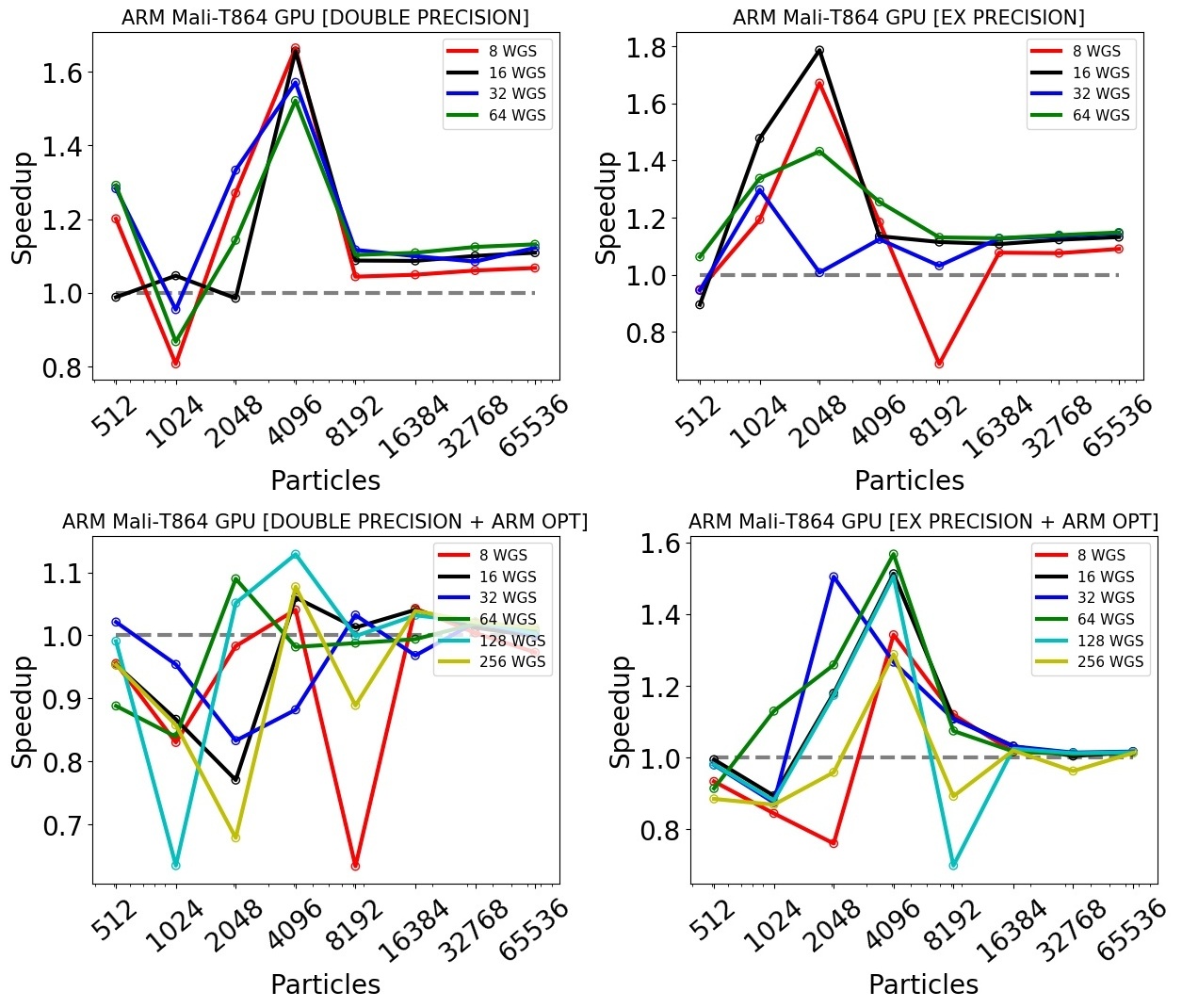}
\caption{GPU speedup for DP-arithmetic (left panels) and EX-arithmetic (right-panels) varying the OpenCL work-group size as a function of the number of particles. Speedup is normalized by the time to solution with work-group of size 4. Top panels for GPGPU kernel code and bottom panels for ARM-GPU-optimized kernel code.}
\label{fig:device_time_D_EX_ARM}
\end{figure}

\begin{figure}[!ht]
\centering\includegraphics[width=\linewidth]{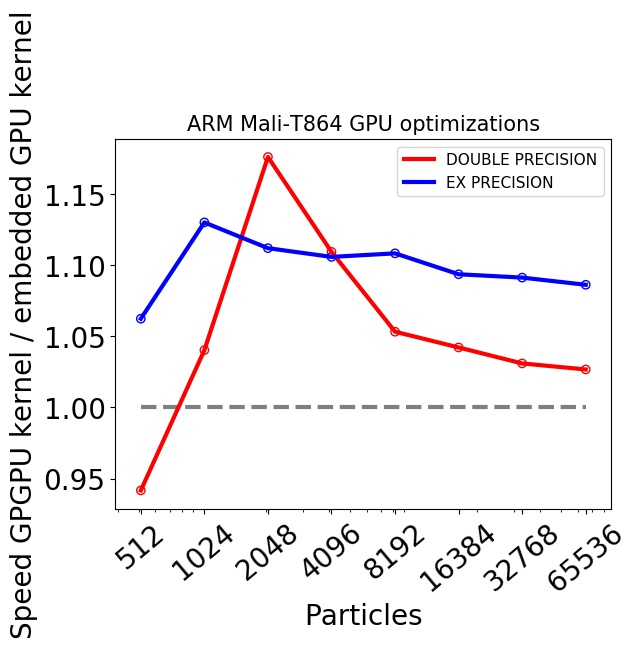}
\caption{Impact of ARM-GPU-optimizations on time to solution for Mali-T864 GPU as a function of the number of particles. Red line for DP-arithmetic and blue-line for EX-arithmetic.}
\label{fig:device_time_ARM_opt}
\end{figure}

The impact of the work-group size on the ARM Mali-T864 GPU performance is studied. Figure~\ref{fig:device_time_D_EX_ARM} shows the speedup achieved varying the OpenCL work-group size for GPGPU kernel code (top panels) and embedded-GPU-optimized kernel code (bottom panels) as a function of the number of particles. The speedup is normalized by the time to solution obtained with work-group size of four.
Kernel execution times on the ARM-GPU have been obtained by means of OpenCL's built-in profiling functionality, which allows the host to collect runtime information.
It is worth noting that work-group sizes of 128 and 256 cause a failure to execute the GPGPU kernel (top panels of Figure~\ref{fig:device_time_D_EX_ARM}) because of insufficient local memory on the GPU. Only ARM-GPU-optimized version of the kernel, which avoids the usage of local memory, can be run with those work-group sizes (the maximum possible work-group size on ARM Mali-T864 is 256).
Despite ARM recommends for best performance using a work-group size that is between 4 and 64 inclusive, the results show that speedup is not driven by any specific work-group size, regardless the usage of local memory. These findings suggest to let the driver to pick the work-group size it thinks as best (the driver usually selects the work-group size as 64).

The impact of embedded-GPU-optimizations are also quantified. Figure~\ref{fig:device_time_ARM_opt} shows the ratio of the best time to solution achieved by GPGPU kernel code and ARM-GPU-optimized kernel code for both arithmetic. In the case of EX-arithmetic the speedup is approximately 10\%, while adopting DP-arithmetic the speedup is nearing 5\% increasing the number of particles.
These findings reveal that adopting the same optimization strategies as those used for high-performance GPGPU computing might lead to worse performance on embedded GPUs. This is in agreement with what was found by \cite{Maghazeh_2013}, when they tested some non-graphic benchmarks on embedded GPUs.

\subsection{ARM CPU-GPU comparison}

\begin{figure}[!ht]
\centering\includegraphics[width=\linewidth]{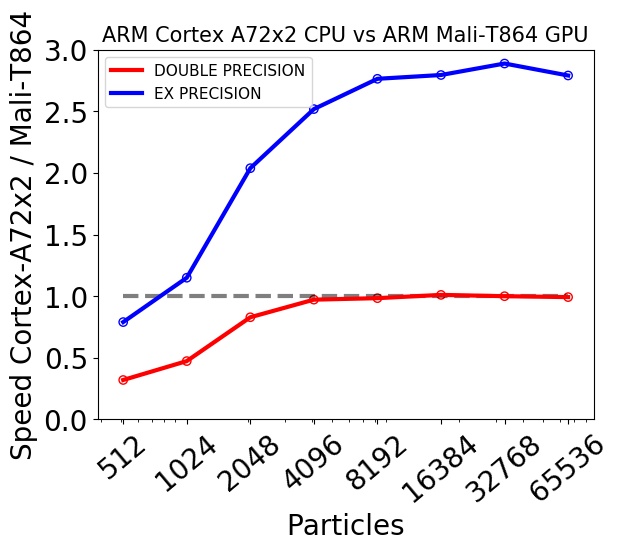}
\caption{Comparison of the time to solution between ARM Cortex-A72x2 CPU and Mali-T864 GPU for both DP-arithmetic (red line) and EX-arithmetic (blue line) as a function of the number of particles.}
\label{fig:CPU_GPU_comparison}
\end{figure}

It is widely accepted that high-end GPGPUs can greatly speedup the solution of the direct $N$-body problem (see section~\ref{section:N_body}).
However in this work we want also to evaluate the performance on low-power embedded GPUs for our kernel.
Figure~\ref{fig:CPU_GPU_comparison} shows the best running time on ARM Cortex-A72x2 as the ratio over the best execution time taken by the ARM-GPU-optimized implementation.
The ARM-GPU-optimized implementation is as fast as the dual-core implementation on the ARM Cortex-A72x2 using DP-arithmetic, as long as the ARM-GPU is kept fed with enough particles, while is almost three times faster using EX-precision.

\section{Future development ad scope}
\label{section:development_scope}

HPC is currently facing, among others, the major technology challenge of the sustainable power consumption. Efficient hardware acceleration is the key to overcome this issue. However, for programmers it is not straightforward to take the mapping decision of a given application to a multi-core CPU or accelerator while optimizing performance.
For this reason, the next step of this research activity is also to quantitatively measure the power-efficiency of \texttt{Hy-Nbody}'s algorithms on ARM-SoC, possibly shedding some light on their suitability for exascale applications.  

On the \texttt{ExaNeSt} prototype the HW acceleration is mainly issued by 
"unconventional"\footnote{Despite FPGAs have been invented in the 1980s, they only start recently to be used in HPC.} FPGA devices, which in comparison to both CPUs and GPUs are more power-efficient (i.e. higher throughput per watt) for different class of applications as shown in the available literature (e.g. \cite{Brodtkorb_2010,Sirowy_2008}). Unlike both CPUs and GPUs, FPGAs do not have any fixed architecture. On the contrary, they provide fine-grain grid of functional units, such as DSP and memory blocks, which can be interconnected to make any desired circuit.
High-level synthesis allows the conversion of an algorithm description in high level languages, e.g. C/C++ or OpenCL, into a digital circuit. However, algorithms have to be modeled for FPGA implementation because the hardware features must be taken into account when attempting to optimize performance. In the case of a CPU or GPU, the programmer tries to achieve the best mapping of a kernel onto a fixed hardware architecture, while for FPGA the aim is to make optimized architecture for that kernel, balancing throughput and resource usage.

The future scope of this research activity is to port \texttt{Hy-Nbody} on FPGA exploiting the \texttt{ExaNeSt} prototype, which is based on Xilinx Zynq Ultrascale+ on SoC FPGA. A state-of-the-art direct $N$-body code, like \texttt{Hy-Nbody}, suitable for real-life astrophysical applications has never been ported on FPGA.
The findings from this work on ARM SoC are fundamental in order to enhance our capabilities to obtain high performance energy-acceleration of kernels for scientific computing on FPGAs.

\section{Conclusions}
\label{section:conclusions}

This research activity has shown that SoC boards can be successfully used to execute a state-of-the-art direct $N$-body code.
The findings reveal that adopting the same optimization strategies as those employed for high-end GPGPUs might not be the best approach on embedded low-power GPUs, because of restricted hardware features. Secondly, embedded GPUs appear to be attractive from a performance perspective as soon as their double-precision compute capability increases. However, the emulated-double-precision approach can be a solution to supply enough power to execute scientific computation and benefit at maximum of SoC devices.

SoC technology will play a fundamental role on future Exascale heterogeneous platforms that will involve millions of specialized parallel compute units. 
Software developer for scientific applications will be forced to design power-efficient algorithms for heterogeneous systems with different devices, and likely with complex memory hierarchies.

Finally, commercial SoC devices such as the Firefly, are
an excellent solution to build low cost testbeds to 
port codes and approach new heterogeneous ARM-based platforms, however they still lack of low latency network that is important when applications are communication bounded more than computing bounded.

\section{Acknowledgments} This work was carried out within the ExaNeSt (FET-HPC) project (grant no. 671553) and the ASTERICS project (grant no. 653477), funded by  the European Union’s Horizon 2020 research and innovation programme.

This research has been made use of IPython \citep{IPython}, Scipy \citep{SciPy}, Numpy \citep{NumPy} and MatPlotLib \citep{MatPlotLib}.


\bibliography{bibtex/mybibfile}
\bibliographystyle{bibtex/splncs_srt}

\end{document}